\documentstyle[12pt,epsf,subeqn,epsfig]{article}

\def\barnui{\:\raisebox{-0.35ex}{$\stackrel{(-)}{\nu}$}\:}
\newcommand{\be}{\begin{equation}}
\newcommand{\ee}{\end{equation}}
\newcommand{\bea}{\begin{eqnarray}}
\newcommand{\eea}{\end{eqnarray}}
\def\nue{{\nu_e}}

\def\numu{{\nu_{\mu}}}

\def\nutau{{\nu_{\tau}}}

\def\lsim{\:\raisebox{-0.5ex}{$\stackrel{\textstyle<}{\sim}$}\:}

\begin{document}

\begin{flushright}
SISSA 18/2003/EP\\
TIFR/TH/03-05\\
hep-ph/0303078
\end{flushright}

\begin{center} 
{\Large \bf 
CP violation and matter effect for a
variable earth density in very long baseline experiments
} \vspace{.5in}  
 
{\bf Biswajoy Brahmachari$^{a,d}$, Sandhya Choubey$^{b}$ and 
Probir Roy$^{c}$\\}  
\vskip .5cm  

$^a$ Theory Group, Saha Institute of Nuclear Physics,\\ 
AF/1 Bidhannagar, Kolkata 700 064, India \\ 

\vskip .5cm    

$^b$ INFN, Sezione di Trieste and
Scuola Internazionale Superiore di Studi Avanzati,\\
I-34014,
Trieste, Italy\\

\vskip .5cm

$^c$ Tata Institute of Fundamental Research, \\ 
Homi Bhaba Road, Mumbai 400 005, India\\

\vskip .5cm

$^d$ Department of Physics, Vidyasagar Evening College,\\
39, Sankar Ghosh Lane, Kolkata 700 006, India\\

\end{center}

\vskip 2cm

\begin{center}
\underbar{Abstract}
\end{center}
\noindent
The perturbative treatment of subdominant oscillation and matter
effect in neutrino beams/superbeams, propagating over long baselines 
and being used to look for CP violation,
is studied here for a general matter density function varying with
distance. New lowest order analytic expressions are given for
different flavour transition and survival probabilities in a general
neutrino mixing basis and a variable earth matter density profile. It
is demonstrated that the matter effect in the muon neutrino
(antineutrino) flavour survival probability vanishes to this order,
provided the depletion, observed for atmospheric muon neutrinos and
antineutrinos at super-Kamiokande, is strictly maximal. This result is
independent of the earth density profile and the distance L between
the source and the detector. In the general variable density case we
show that one cannot separate the matter induced asymmetry from a
genuine CP effect by keeping two detectors at distances $L_1$ and
$L_2$ from the source while maintaining a fixed ratio $L_1/E_1 =
L_2/E_2$. This needs to be done numerically and we estimate the
asymmetry generated by the earth matter effect with particular
density profiles and some chosen parameters for very long baseline 
neutrino oscillation experiments.
 
\newpage

\section{Introduction}
Finding CP violation in the neutrino sector \cite{ref1} is a tantalizing
goal waiting to be attained \cite{ref2} in forthcoming long baseline
experiments with neutrino beams/superbeams and future ones at
neutrino factories. The discovery and quantitative measurement
of such an effect will not only open a new window for physics
beyond the standard model, but {\it may} provide an insight into
leptogenesis \cite{ref3}. From solar neutrino studies and the reactor
experiment by KAMLAND, we already know
\cite{ref4} the squared mass difference and the angle of mixing
between concerned mass eigenstate pair $\nu_{1,2}$ to be
$\Delta m^2_{21} \equiv m^2_2-m^2_1 \sim 7\times 10^{-5}~eV^2$ 
and $\sin^2\theta_{12} \sim 0.3$ respectively 
(see \cite{klpaper} and references therein). Similarly,
atmospheric neutrino studies \cite{ref5} have revealed the
corresponding parameters for the other mass eigenstate pair 
$\nu_{2,3}$ to be $|\Delta m^2_{32}| \equiv |m^2_3-m^2_2| \sim 
3 \times 10^{-3}~eV^2$ and $\sin^2\theta_{23} \sim 0.5$. Other 
reactor experiments \cite{ref6} have, however, shown that the third 
possible mixing angle $\theta_{13}$ is very small, consistent with zero.
These mysterious numbers have increased the importance of the task
of getting a quantitative handle on CP violation likely to be
associated with the MNS neutrino mixing matrix. This is sought
\cite{ref1,ref2} 
to be achieved by measuring the difference between neutrino and
antineutrino beams in the muonic to electronic flavour transition
probability which can be obtained from the general expression
\be
\Delta P_{\beta \alpha}(L,E)=
P[\nu_\alpha(0) \rightarrow \nu_{\beta}(L)]-
P[\overline{\nu}_\alpha(0) \rightarrow 
\overline{\nu}_{\beta}(L)] \label{asy}
\ee
over a large distance $L$, $E$ being the beam energy and $\alpha$, 
$\beta$ being the flavour indices.
If $|\Delta m^2_{32}| L/E$ is chosen to be
${\cal O} (1)$, the effect will be driven
dominantly by $\nu_\mu - \nu_\tau$ oscillation. 

A major practical problem, associated with the above task, is the
occurrence of the matter effect in neutrino flavour transitions
studied in long baseline experiments. These contribute to 
$\Delta P_{\beta \alpha}(L)$ and induce a ``fake CP violation''. The
latter needs to be  
filtered out, leaving only the genuine CP violation part. 
However, the matter effect is not merely a background but deserves to be
studied in its own right. In particular, it can yield valuable  
information on the sign of
$\Delta m^2_{32}$, on the MSW resonant enhancement of neutrino oscillations
and on the mixing between neutrinos of the first and the 
third generations. It is important
therefore to be able to compute the matter effect relevant to any given
experiment looking for CP violation.
Several studies
\cite{ref2,ref8,Freund:1999gy,ref9} have been conducted to this end. 
Specifically, we shall use the formalism of Arafune et al. \cite{ref9}
-- developed to treat the matter effect as well as the subdominant
oscillation driven by 
$\Delta m^2_{21}$ and $\theta_{12}$ perturbatively to the lowest order
relative to the dominant oscillation driven by $|\Delta m^2_{32}|$
and $\theta_{23}$. Thus the evolution matrix element 
$S_{\beta \alpha} (x) \equiv <\nu_\beta (x)|\nu_\alpha(0) >$,
$\alpha,\beta$ being flavour indices, could be calculated to first
order in $\Delta m^2_{21}$ and $a$, where 
\be
a(x,E)= 2 \sqrt{2} G_F N_e E =7.56 \times 10^{-5}
~{\rm eV^2}~{\rm E \over GeV}~{\rm \rho(x) \over gms/cc},
\ee
with $N_e$ being the electron density and $\rho(x)$
the mass density of the earth, expressed as a function of the path
length $x$ of the beam.  The parameter $\rho$ was assumed in Ref. \cite{ref9}
to be spatially uniform\footnote{For instance, one could assume 
\cite{Freund:1999vc} a constant
density $\rho$ equal to the average density of the PREM profile
\cite{ref10}.}. This is not always a realistic assumption \cite{ref8},
especially for future very long baseline experiments where
nonuniformity in earth's density profile cannot be neglected.
One possible approach \cite{ref11} is to write the density function
$\rho (x)$, and correspondingly $a(x,E)$, as an average constant plus a
spatially fluctuating part and expand the latter into Fourier modes,
arguing that only the first few modes is important for foreseable
experiments. There is, however, a large uncertainly 
\cite{ref12,Shan:2001br,Shan:2003vh} in
the seismological knowledge of the latter; indeed, this feeds
back into the average density parameter. Under the circumstances, it
is worthwhile making --- within the lowest order perturbative
framework --- general statements for an arbitrarily varying $a(x,E)$
that can be checked by experimental measurements. This is our aim
here. Our approach is basically analytic and the price to pay is to
use lowest order perturbation theory.  As will be pointed out in our
section on numerical studies, a lowest order perturbative result is
often invalid in very long baseline neutrino experiments.  There are,
however, sizable domains in the $(L,E)$ plane where it should be
reliable.  At least, the importance of the effects considered is
clearly brought out by our analytic considerations.

In this paper we reexamine the treatment of Ref. \cite{ref9}, assuming
an arbitrary spatial dependence in $\rho(x)$ and hence $a(x,E)$,
consistent with the approximation $|a| << |\Delta m^2_{32}|$. We give
extended versions of the formulae of Arafune et al, accommodating such
a nonuniform $\rho(x)$ and in a general neutrino mixing basis. Even
with such an arbitrary earth density profile, we find that the matter
effect in the muon neutrino (antineutrino) flavour survival
probability $P[\barnui_{\!\!\!\!\!\mu}(0) \rightarrow
\barnui_{\!\!\!\!\!\mu}(L)]$ vanishes if the flavour conversion of
atmospheric $\nu_\mu$ and $\bar\nu_\mu$, as observed in super-K, is truly
maximal. However, the simple methods, proposed in Ref. \cite{ref9} to
filter out the matter effect terms, are specific to the uniform earth
density assumption and do not extend to the variable density case in
the lowest order of perturbation theory. We
show that, in general, it is not possible to separate the `fake'
matter induced asymmetry from the genuine CP effect by keeping two
detectors at distances $L_1$ and $L_2$ from the source while keeping
the ratio $L_1/E_1 = L_2/E_2 = L/E$ fixed and taking a linear combination
of $P[\nu_\mu (0) \to \nu_\beta (L)]$ and $P[\bar\nu_\mu (0) \to
\bar\nu_\beta (L)]$, as proposed for a constant matter density profile
to the lowest perturbative order in subdominant oscillation and matter
effects by Arafune et al. This separation has to be done numerically and in
the last part we compute the
asymmetries generated by the earth's matter effect for sample earth
matter density profiles keeping specific experimental possibilities in
mind. 

This paper is organized as follows. Section 2 contains an extension of
the lowest order perturbative calculation of the evolution operator to
the variable density case. In Section 3 we calculate the muon
(anti)-neutrino flavour survival probability
$P[\barnui_{\!\!\!\!\!\mu} (0) \to
\barnui_{\!\!\!\!\!\mu} (L)]$. This calculation is extended to the general
flavour transition probability $P [\barnui_{\!\!\!\!\!\alpha} (0) \to
\barnui_{\!\!\!\!\!\beta} (L)]$ in Section 4. Our numerical studies are
presented in Section 5. Section 6 contains a summary of our results. 

\section{Evolution operator formalism}

The formulae for the amplitude and
probability of the transition of $\nu_\alpha(0)$ to $\nu_\beta(L)$ in
vacuum are\footnote{We shall use $L$ to denote the baseline with 
respect to a measurement, $x$ to
denote any
intermediate length and $s$ for a dummy variable in any integration being
performed upto $x$.}
\subequations 
\bea
A(\nu_\alpha (0) \rightarrow \nu_\beta (L)) &=& < \nu_\beta(L)|\nu_\alpha(0) >
= \sum_{i}U^*_{\alpha i} e^{\displaystyle{-m^2_i L \over 2 E}} U_{\beta i}, \\
P(\nu_\alpha (0) \rightarrow \nu_\beta (L)) 
&=& \delta_{\alpha \beta} -4 \sum_{j>i}
Re(U_{\alpha i} U^*_{\beta i} U^*_{\alpha j} U_{\beta j})
\sin^2 {\Delta m^2_{ij} L \over 4E} \nonumber\\
&& + 2 \sum_{j >i}Im(U_{\alpha i} U^*_{\beta i} U^*_{\alpha j}
U_{\beta j}) \sin {\Delta m^2_{ij} L \over 2E},
\eea
\endsubequations
where $\Delta m^2_{ij}=m^2_i-m^2_j$ and the neutrino mixing matrix $U$
is defined in vacuum by $|\nu_\alpha > = U_{\alpha i} |\nu_i >, \alpha$
and $i$ being flavour and mass eigenstate indices respectively. More
generally, when matter is present, one can decompose the Hamiltonian
as 
\be
H=H_0+H^\prime ,
\ee
where $H_0$ is the unperturbed part containing $\Delta m^2_{32}$ while
$H^\prime$ is the perturbation involving $\Delta m^2_{21}$ and
$a(x,E)$. 

%Let us motivate the need for the evolution operator formalism. 
%Only if $H^\prime$ is x-independent can equations (3) be directly extended
%\cite{ref13} to include the matter effect by the replacements 
%$m^2_i \rightarrow M^2_i$, $\Delta m^2_{ij} \rightarrow \Delta M^2_{ij}$ and 
%$U \rightarrow {\ut}$ where $M^2_i ~(\Delta M^2_i)$ are matter
%dependent squared masses (mass differences) of the neutrinos and ${\ut}$
%is the matter dependent neutrino mixing matrix. One can do this since now
%\be
%|\nu(x) > = e^{-ixH} | \nu(0) >
%\ee
%for any $x$. In fact, (3a) and (3b) can then be written for any intermediate
%length $x$ replacing $L$. However, when the perturbation $H^\prime$ is
%x-dependent (and the only source of such a dependence is spatial
%variation in the earth's matter density function), such an extension 
%breaks down since the extended versions of (3) are no longer true. Instead,
%we have 
%\be
%|\nu(x) > =T e^{\displaystyle{-i \int_0^x ds H(s)}} | \nu(o) > .
%\ee
%Since the RHS of (5) is not $e^{-i x H(x)} | \nu(0) >$, 
%\be
%A( \nu_\alpha(0) \rightarrow \nu_\beta(x)) \neq  
%\sum_{i} {\ut}^*_{\alpha i} 
%e^{\displaystyle{- {M^2_{i} x \over 2E}}} {\ut}_{\beta i}
%\ee
%for every $x$. In fact, every bit of $dx$ has to be integrated 
%starting from $x=0$ and one is obliged in the variable
%density case to use the evolution operator formalism. 

In the case of an $x$-dependent $H^\prime$, we cannot use the
replacement procedure of Ref. [18], but need to use the evolution
operator formalism.  Thus we write
\be
|\nu_\beta (x)>  =  S_{\beta\alpha} (x) | \nu_\beta (0) > ,
\ee
%with
%\subequations
%\begin{eqnarray}
%S(x)  &=&  T e^{\displaystyle{-i \int^x_0 dy H(y)}}, \\
%S(0)  &=&  1 .
%\end{eqnarray}
%\endsubequations
where the operator $S$ obeys the evolution equation
\be 
i {dS \over dx}  =  H_0 S(x) + H'(x) S (x) ,
\label{HO}
\ee
with
\subequations
\bea
H_0 &=& {1 \over 2E} 
{U} \pmatrix{0 & 0 & 0 \cr
                 0 & 0 & 0 \cr  
                 0 & 0 & \Delta m^2_{32}} {U}^\dagger,\\
H^\prime (x) &=& {1 \over 2E}\left[ 
{U} \pmatrix{0 & 0 & 0 \cr
                 0 & \Delta m^2_{21} & 0 \cr  
                 0 & 0 & 0} {U}^\dagger + 
                 \pmatrix{ a(x,E) &0 &0 \cr
                           0 & 0 & 0 \cr
                           0 & 0 & 0 } \right] .
\eea
\endsubequations
%$a(x,E)$ being given by (2).
%Now we define
%\subequations
%\begin{eqnarray}
%\Omega(x) &\equiv&  e^{ixH_0} S(x) , \\
%\Omega(0) &=& 1 .
%\end{eqnarray}
%\endsubequations
%Also, let us define 
%\be
%{\cal H}^\prime(x) \equiv e^{ixH_0} H^\prime(x) e^{-ixH_0}.
%\ee
%Then the following differential equation obtains 
%for $\Omega(x)$:
%\be
%i{d \Omega(x) \over dx} = {\cal H}^\prime (x) \Omega(x) ,
%\ee
%with the solution 
%\subequations
%\bea
%\Omega(x) &= & Te^{\displaystyle{-i \int^x_0 ds {\cal H}^\prime(s)}} \\ 
%& \simeq&  1-i \int_0^x ds {\cal H}^\prime(s)
%\eea
%\endsubequations
To the lowest order of perturbation, (\ref{HO}) can be solved by \cite{ref9} 
\be
S(x)  \simeq  S^0(x) + S^\prime(x) , 
\ee
with
\subequations
\bea
S^0(x) &=& e^{\displaystyle{-ixH_0}} ,\\
S^\prime(x) &=& e^{\displaystyle{-ixH_0}} (-i) \int^x_0 ds\ e^{isH^0}
H'(s) e^{-isH^0} .
\eea
\endsubequations

We turn now to matrix elements in the flavour basis and work in the
approximations $\Delta m^2_{21} \ll |\Delta m^2_{32}|$ and $|a(S,E)| \ll
|\Delta m^2_{32}|$.  First, it is convenient to define
\be
g(v) \equiv {1\over4} \Delta m^2_{31} v.
\ee
Then we can explicitly write
\bea
S^0 (x)_{\beta\alpha} &=& \delta_{\beta\alpha} + U_{\beta 3}
U^\star_{\alpha 3} \left[e^{-ig(2x/E)}-1\right] \nonumber \\[2mm] 
&=& \delta_{\beta\alpha} - 2i U_{\beta 3} U^\star_{\alpha 3}
e^{-ig(x/E)} \sin [g(x/E)] \nonumber \\[2mm] 
&\equiv& A_{\beta\alpha} (x/E)
\eea
for the unperturbed part.  The lowest order expression for the
perturbed part 
\[
S'(x)_{\beta\alpha} = -i \int^x_0 ds
\left[e^{-i(x-s)H_0}\right]_{\beta\gamma}
\left[H'(s)\right]_{\gamma\delta}
\left[e^{-sH_0}\right]_{\delta\alpha}
\]
can be rewritten, on using (7a), as
\bea
S'(x)_{\beta\alpha} &=& -i \int^x_0 ds U_{\beta i} \exp\left[-i{\rm diag}
\left\{0,0,g\left(2(x-s)/E)\right)\right\}\right]_{ii}
U^\star_{\gamma i} \nonumber \\[2mm] 
&& \left[H'(s)\right]_{\gamma\delta} U_{\delta j} \exp\left[-i {\rm diag}
\left\{0,0,g(2s/E)\right\}\right]_{jj}
U^\star_{\alpha j}.
\eea

In the RHS of (12), $H^\prime(s)$ has two additive parts, one
constant and one depending on $s$:
\be
H^\prime(s) = H^1+H^a(s) , 
\ee
with
\subequations
\bea
H^1 &= &{1 \over 2E} {U} 
\pmatrix{0 & 0 & 0 \cr 0 & \Delta m^2_{21}& 0 \cr 0 &0 &0}{U}^\dagger , \\
H^a &= &{1 \over 2E} 
\pmatrix{a(s,E) & 0 & 0 \cr 0 & 0 & 0 \cr 0 & 0 & 0}.
\eea
\endsubequations
Correspondingly, we can take  
\be
S^\prime(x)_{\beta \alpha}
=S^1(x)_{\beta \alpha}+S^a(x)_{\beta \alpha} ,
\ee
where $H^1$ contributes to $S^1$ and $H^a$ to $S^a$. Since $H^1$
does not depend on s, we may write 
\bea
S^1(x)_{\beta\alpha}\!\!\!\!&=&\!\!\!\! -i~{U}_{\beta i}~ 
{U}^*_{\gamma i}~(H^1)_{\gamma \delta}~
{U}_{\delta j} {U}^*_{\alpha j} ~~ . \nonumber \\ 
&& \int^x_0 \!\!\!ds \exp \left[- i{\rm diag}\left\{0,0,g(2(x-s)/E)
\right\}\right]_{ii}  \exp 
\left[- i{\rm diag}\left\{0,0, g(2s/E)\right\}\right]_{jj} 
\nonumber \\
&=& -i {\Delta m^2_{21} \over 2E}  U_{\beta 2}~U^\ast_{\alpha 2} \equiv -i 
B_{\beta \alpha} (x/E) \Delta m^2_{21}, 
\label{s1ba} 
\eea
where we have used the identity 
\bea
{U}^*_{\gamma i}~[H^1]_{\gamma \delta}~{U}_{\delta j}
= { \Delta m^2_{21} \over 2E}~\delta_{i2}~\delta_{j2}.\nonumber   
\eea

Turning to $S^a(x)_{\beta \alpha}$, it is convenient to use the result
\be
{U}^*_{\gamma i}~[H^a(s)]_{\gamma \delta}~{U}_{\delta j}
={a(s,E) \over 2E} {U}^*_{1i}~{U}_{1j}.\label{tt}
\ee
The employment of (\ref{tt}) enables us to write $S^a (x)_{\beta
\alpha}$ as   
\bea
S^a_{\beta \alpha}(x, E) &=& - {i \over 2E} ~{U}_{\beta i}
{U}^*_{\alpha j} {U}^*_{1 i} {U}_{1 j} \int^x_0 ds~ 
\exp \left[- i g \left\{\frac{(2x - 2s) \delta_{i3} + 2s \delta_{j3}} {E}\right\}\right] a(s,E), \nonumber \\
&=& - {i \over 2E} ~{U}_{\beta i} {U}^*_{\alpha j} {U}^*_{1 i}
{U}_{1 j}~\Gamma^a_{ij} (x, E) , \label{sa1} 
\eea
with  
\bea
\!\!\!\!\Gamma^a_{ij} (x, E)  &=& \delta_{i3}~\delta_{j3}~e^{\displaystyle{
-i g (2x/E)}} \int^x_0 ds~a(s,E)
+ ~~(1-\delta_{i3})~(1-\delta_{j3})~\int^x_0 ds~a(s,E) \nonumber\\
&+& !\!\!\!\!\!(1-\delta_{i3})\delta_{j3} \int^x_0 ds~a(s,E)
e^{\displaystyle{-i g (2s/E)}}
+ \delta_{i3}(1-\delta_{j3})
\int^x_0 ds ~a(s,E)~e^{\displaystyle{-i g (2(x-s)/E)}}. \nonumber \\
&&
\eea
On using (19) in (18), we obtain
\bea
S^a_{\beta \alpha} (x,E) &=& - {i \over 2E} ~ \delta_{\beta 1} \delta_{1 \alpha} 
\int^x_0 ds ~ a (S,E) \nonumber \\
&& - {1 \over E} \delta_{1 \alpha} {U}_{\beta 3}
{U}^*_{13} \int^x_0 ds~a (s,E)~ e^{-i[g (x/E) - g(s/E)]} \sin [g (x/E) - g(s/E)] 
\nonumber  \\
&& - {1 \over E} \delta_{\beta 1} U^*_{\alpha 3} U_{13} \int^x_0 ds~a (s,E)~ 
e^{-ig(s/E)} \sin [g(s/E)] \nonumber \\
&& - {i \over 2E} U^*_{\alpha 3} U_{\beta 3} |U_{13}|^2 e^{-ig (x/E)}~ R(x,E) 
\nonumber \\
& \equiv & - i G^a_{\beta \alpha} (x,E).  
\label{sa}
\eea
In (\ref{sa}) we have introduced the real function $R(x,E)$: 
\bea
R (x,E) \!\!\!&\equiv&\!\!\! e^{-i g(x/E)}
\int^x_0 ds~a(s,E) \{ 1 - e^{i g(2s/E)} \} +
e^{i g (x/E)} \int^x_0 ds~a(s,E) \{ 1 - e^{-i g(2s/E)})\}, \nonumber \\
\!\!\!&\equiv&\!\!\! 2 \int^x_0 ds ~ a (s,E) \left[\cos g (x/E) - \cos \left\{g (x/E) - g(2s/E) \right\}\right].   \label{m}
\label{mr}
\eea
Note that $R(x,E)$ becomes a function of $x/E$ for a constant 
$\rho(x)$, not otherwise.

Finally, the flavour matrix element of the evolution operator $S_{\beta \alpha} = S^0_{\beta\alpha} + S^1_{\beta\alpha} + S^a_{\beta\alpha}$ can be expressed as 
\be
S_{\beta\alpha} (x,E) = A_{\beta \alpha} (x/E) - i B_{\beta\alpha} (x/E)
\Delta m^{2}_{21} - i G^a_{\beta \alpha} (x,E) ,  \label{s}
\ee
where $A_{\beta\alpha}(x/E), B _{\beta\alpha} (x/E)$ and $G^a_{\beta \alpha} (x,E)$ are given by (11), (\ref{s1ba}) and
(\ref{sa}) 
respectively. 
We can now calculate the transition probability $P_{\beta\alpha}$ to
the lowest order in $a$ and $\Delta m^2_{21}$ in the approximations
$\Delta m^2_{21} \ll |\Delta m^2_{31}|, |a(s,E)| \ll |\Delta m^2_{31}|$
stated already.
We obtain
\bea
P_{\beta\alpha} &=& P^0_{\beta\alpha} + P^\alpha_{\beta\alpha}, 
 \label{p} 
\eea
with 
\be
P^0_{\beta\alpha} = A^*_{\beta\alpha} A_{\beta \alpha} + 2 Im 
(A^*_{\beta\alpha} B_{\beta \alpha}),
\ee
\be
P^a_{\beta\alpha} =  2 Im 
(A^*_{\beta\alpha} G^a_{\beta \alpha}),
\ee

\section{The flavour survival probability 
$\nu_{\mu} (0) \rightarrow \nu_{\mu} (L) $}
%$\barnui_{\!\!\!\!\!\mu} \rightarrow \barnui_{\!\!\!\!\!\mu}$}

Using (11), (\ref{s1ba}), (20) and choosing 
$\alpha=\beta=\mu$, we have 
\subequations
\bea
A_{\mu\mu} (x/E) &=&
1-2i~|U_{\mu 3}|^2 e^{-ig(x/E)} \sin[g (x/E)] , \\
B_{\mu\mu} (x/E) &=& |U_{\mu 2}|^2 { x \over 2 E} , \\
G^a_{\mu\mu}(x,E) &=&  {|U_{e3}|^2|U_{\mu 3}|^2 \over 2 E} e^{-ig(x/E)} R(x,E), 
\eea
with $R(x,E)$ as defined in (\ref{mr}).
\endsubequations
In the matter free case, $a(s,E)=0$ and replacing $x$ by $L$ we obtain 
\bea
P^0_{\mu \mu} (L/E) &=& 1- 4|U_{\mu 3}|^2\sin^2[g (L/E)] \nonumber \\
&+&
4|U_{\mu 3}|^4\sin^2[g (L/E)] + {L \over E} 
|U_{\mu 2}|^2|U_{\mu 3}|^2\sin [2 g (L/E)]
\Delta m^2_{21}. \label{37}  
\eea
If we consider this transition to be overwhelmingly driven by a two flavour
oscillation, as done in the super-K analysis \cite{ref5}, we can
ignore the third RHS term in (\ref{37}) to see that the depletion
$1-P^0_{\mu\mu} (L,E)$, i.e. the flavour transformation and hence mixing
for a muonic neutrino or antineutrino, would be maximal for $|U_{\mu
3}| = 1/\sqrt{2}$. Next, we consider propagation in matter and keep
$a$ and hence $M(L)$. Then we have 
\bea
P_{\mu \mu} (L,E) &=& P^0_{\mu \mu} (L/E) + P^a_{\mu \mu} (L,E) ,
\eea
where 
\be
P^a_{\mu \mu} (L,E) = {|U_{e3}|^2 |U_{\mu 3}|^2 \over E}~(2|U_{\mu 3}|^2
-1)~\sin[g (L/E)]~R(L,E) . 
\ee
The asymmetry 
$\Delta P_{\mu \mu} (L,E)$, defined in (\ref{asy}), is thus 
\bea
\Delta P_{\mu\mu} (L,E) &=& {|U_{e3}|^2 |U_{\mu 3}|^2 \over E} (2|U_{\mu
3}|^2 -1) ~\sin[g(L/E)]~[R^+(L,E)-R^-(L,E)] \nonumber \\
&=& {2|U_{e3}|^2 |U_{\mu 3}|^2 \over E} (2|U_{\mu 3}|^2 -1) \sin [g(L/E)]
R^+(L,E) .\label{40} 
\eea
In (\ref{40}) $R^{\pm}(L,E)$ is obtained from (21) by using
$a(s,E) \equiv \pm |a(s,E)|$ respectively so that $R^-(L,E) = - R^+ (L,E)$. 
Thus we see that the {\it matter effect contribution to the muonic
flavour survival probability} $P_{\mu\mu} (x)$ {\it vanishes if}
$|U_{\mu 3}| = 1/\sqrt{2}$. Moreover, even in matter and to the lowest
order of perturbation, any nonzero
asymmetry $\Delta P_{\mu\mu}$, detected from the muon (anti) neutrino
flavour survival probabilities $P (\barnui_{\!\!\!\!\!\mu}
\rightarrow \barnui_{\!\!\!\!\!\mu})$ will signal a deviation 
from the condition for the strictly maximal mixing of atmospheric
neutrinos at super-K. This can be used in future to sensitively probe
any deviation of $|U_{\mu 3}|$ from its maximal value $1/\sqrt{2}$. 

\section{General oscillation probability in matter: 
$\nu_\alpha \rightarrow \nu_\beta $}

From (24) and (26a,b) we have
\bea
P^0_{\beta\alpha} (L/E) &\simeq& \delta_{\beta\alpha} \left[1-4 |U_{\alpha 3}|^2 \sin^2
[g(L/E)]\right] + 4 |U_{\alpha 3}|^2 |U_{\beta 3}|^2 \sin^2 [g(L/E)]
\nonumber \\
&& + {L \Delta m^2_{21} \over E} \bigg[Re~(U^*_{\alpha 3} U_{\beta
3} U_{\alpha 2} U^*_{\beta 2}) \sin [2g(L/E)] \nonumber \\
&& - 2~Im~(U^*_{\alpha 3}
U_{\beta 3} U_{\alpha 2} U^*_{\beta 2}) \sin^2 [g(L/E)]\bigg] ,
\label{44}
\eea
while (25) and (26a,c) lead us to the expression
\bea
P^a_{\beta\alpha} (L,E) &\simeq& 2~Im \left[A^*_{\beta\alpha} (L/E)
G^a_{\beta\alpha} (L,E)\right] \nonumber \\
&=& \delta_{\beta\alpha} \bigg\{{1\over E} \delta_{1\alpha} |U_{\alpha
3}|^2 \sin^2 [g(L/E)] \int^L_0 ds~a(s,E) \sin [g(L/E)-g(s/E)]\nonumber \\ 
&& - {1\over E}
|U_{\alpha 3}|^2 |U_{e3}|^2 \sin [g(L/E)] R(L,E)\bigg\} 
\nonumber \\
&& + {1\over E} |U_{\alpha 3}|^2 |U_{\beta 3}|^2 \bigg\{2 |U_{e3}|^2 
(\delta_{1\alpha} + \delta_{\beta 1}) \sin [g(L/E)] \bigg\} R(L,E), 
\label{45}
\eea
with $R(L,E)$ substituted from (21). 

\eject

We can now discuss two distinct cases.

\noindent \underbar{Case 1: $\alpha=\beta$}

\begin{itemize}
\item
$\alpha = \beta \neq e$ 

In this case the matter independent and matter dependent transition
probabilities, cf.

\subequations
\bea
\!\!\!\!P^0_{\alpha \alpha} (L/E)\!\!\!\! &=& \!\!\!\!1 \!\!- \!\!4 |U_{\alpha
  3}|^2 (1- |U_{\alpha 
3}|^2) \sin^2 [g (L/E)] + {\Delta m^2_{21} L \over E} |U_{\alpha 2}|^2
|U_{\alpha 3}|^2 \sin [2g (L/E)] , \label{46a} \\
\!\!\!\!P^a_{\alpha \alpha} (L,E) \!\!\!\!&=&\!\!\!\! {1 \over E} |U_{\alpha 3}|^2 |U_{e3}|^2
(2 |U_{\alpha 3}|^2-1) \sin [g (L/E)] R(L,E) .\label{46b}
\eea
\endsubequations

\item
$\alpha = \beta = e$ 

The expression for $P^0_{ee} (L)$ is the same as the RHS of
(\ref{46a}) with $\alpha = e$, but the matter dependent part is 
\bea
P^a_{ee} (L,E) &=& {1\over E} |U_{e3}|^2 \sin^2 [g(L/E)] \int^L_0 ds~a(s,E)
\sin [g(L/E) \nonumber \\&&
-k(s/E)/2] - {1\over E} |U_{e3}|^4 \sin [g(L/E)] R(L,E)
 \nonumber \\
 && 
+ {2\over E} |U_{e3}|^4 (|U_{e3}|^2 -1) \sin [g(L/E)] R (L,E). 
\eea
\end{itemize}
%Note that, for $a(s) =$ constant, $R$ as well as $P^a_{ee}$ are
%functions of $x/E$ -- not otherwise.

\noindent \underbar{Case 2: $\alpha \neq \beta$}

Now the matter independent transition probability is given from
(\ref{44}) by 
\bea
\!\!\!\!\!\!P^0_{\beta\alpha} (L/E) \!\!\!\!&=& \!\!\!\!4|U_{\beta 3}|^2 
|U_{\alpha 3}|^2 \sin^2 [g(L/E)] 
\!\!-\!\! {2\Delta m^2_{21} L \over E} \left\{Im\xi ~\sin^2 [g(L/E)] 
\!\!- \!\!{1\over 2} Re\xi ~\sin [2g (L/E)]\right\},  \label{p01}
\eea
with 
\be
\xi \equiv U^*_{\beta 2} U_{\beta 3} U_{\alpha 2} U^*_{\alpha 3} .
\ee
For the matter dependent part, (\ref{45}) leads to three cases. 
\begin{itemize}
\item 
$\alpha \neq e, \ \beta \neq e$
\be
P_{\beta \alpha}^a (L,E) = {2 \over E} {|U_{\alpha 3}|^2 |U_{\beta 3}|^2 
|U_{e3}|^2} \sin[g(L/E)] R(L,E). \label{pa1} ,
\ee 
\item 
$\alpha \neq e, \ \beta = e$
\be
P_{e \alpha}^a (L,E) = {1\over E} |U_{\alpha 3}|^2 |U_{e3}|^2
(2|U_{e3}|^2 -1) \sin [g(L/E)] R(L,E) .  
\ee
\item 
$\alpha = e, \ \beta \neq e$
\be
P_{\beta e}^a (L,E) = {1\over E} |U_{e3}|^2 |U_{\beta 3}|^2
(2|U_{e3}|^2 -1) \sin [g(L/E)] R(L,E) .
\ee
\end{itemize}
%Once again, only for a constant $a(x)$, are $R$ and hence RP^e$
%functions of $x/E$, not othersie. 

Both the CP violating part, proportional to $Im\xi$ in (\ref{p01}),
and the matter dependent part $P^a_{\beta\alpha}$ change sign when one
goes from neutrinos and antineutrinos. However, the former is a
function of $x/E$, while the latter involves both $x/E$ and $x$ in an
unknown way for a general earth matter density function $\rho(x) = (2
\sqrt{2} G_F N_eE)^{-1} a(x,E)$. Only for $\rho(x)$ = constant, can all the
transition probabilities become functions of $L/E$ and the
matter dependent part can be eliminated, to the lowest order of
perturbation, by [9] taking $L_1 (L_2 - L_1)^{-1} [\Delta
  P_{\beta\alpha} (L_2) - \Delta P_{\beta\alpha} 
(L_1)]_{L/E~fixed}$, but this procedure is in general invalid for a
spatially varying $a(x,E)$. 

\section{Numerical estimate of matter induced asymmetry}

In this section we present some numerical results for the 
matter contribution to the transition 
probabilities using the formalism developed in this paper.
We choose realistic neutrino beam energies and 
both realistic and notional detector baselines and discuss the conditions 
for the validity of our 
perturbative calculation.
%In Table \ref{expt} we present the 
%mean energies and approximate baselines for the proposed experiments
%involving Neutrino Factories and Superbeams.
For numerical studies we have used the 
Preliminary Reference earth Model(PREM) \cite{ref10} and 
another earth model -- ak135-F \cite{ak135f}.
For the validity of perturbation theory for all intermediate
values of $s$ we must satisfy the conditions 
\begin{eqnarray}
\frac{a(s,E)s}{4E} &<<& \frac{\Delta m^2_{31} s}{4E} \label{eq01} \\
{a(s,E) s \over 2 E} << 1 ;&&
{\Delta m^2_{21} s \over 2E} << 1 \label{eq02}\\
\sin^2 \left( \frac{\Delta m^2_{21} s} {4E}\right ) &<<& 
\sin^2 \left (\frac{\Delta m^2_{31} s}{4E} \right ),\label{eq03} 
\end{eqnarray}
where $0 \leq s \leq L$.
Note that conditions in eqns (\ref{eq02}) 
are required for the expansion of $S (x)$ in Section 2.
Neutrino oscillation experiments either look for disappearance of 
the initial neutrino beam or for the appearance of a different 
flavour in the final neutrino beam at the detector. Both these effects are 
maximal when $\sin^2(\Delta m_{31}^2 L/4E)\sim 1$ corresponding 
to a ``peak'' in the transition probability. 
The Super-Kamiokande 
atmospheric neutrino data demands a $\Delta m_{31}^2 \sim 3\times 10^{-3}$ 
eV$^2$ \cite{ref5} while the mass squared difference associated with the 
solar neutrino oscillation has now been confirmed to be around 
$\Delta m_{21}^2 \sim 7\times 10^{-5}$ eV$^2$ 
by the KamLAND experiment \cite{klpaper}. Thus for all realistic 
experimental scenarios where the beam energy is tuned to an oscillation 
maximum for $\Delta m_{31}^2$, oscillations 
driven by the solar scale is always subdominant compared to those 
driven by the atmospheric scale, i.e.
the condition (\ref{eq03}) and the second of conditions (\ref{eq02}) are
satisfied. 
For the validity of our approximation 
(\ref{eq03}) 
we will therefore always confine ourselves to values of 
\begin{eqnarray}
{ L/Km \over E/GeV} &\sim& O(10^2-10^3). 
\end{eqnarray}

\begin{figure}[t]
\centerline{\epsfig{figure=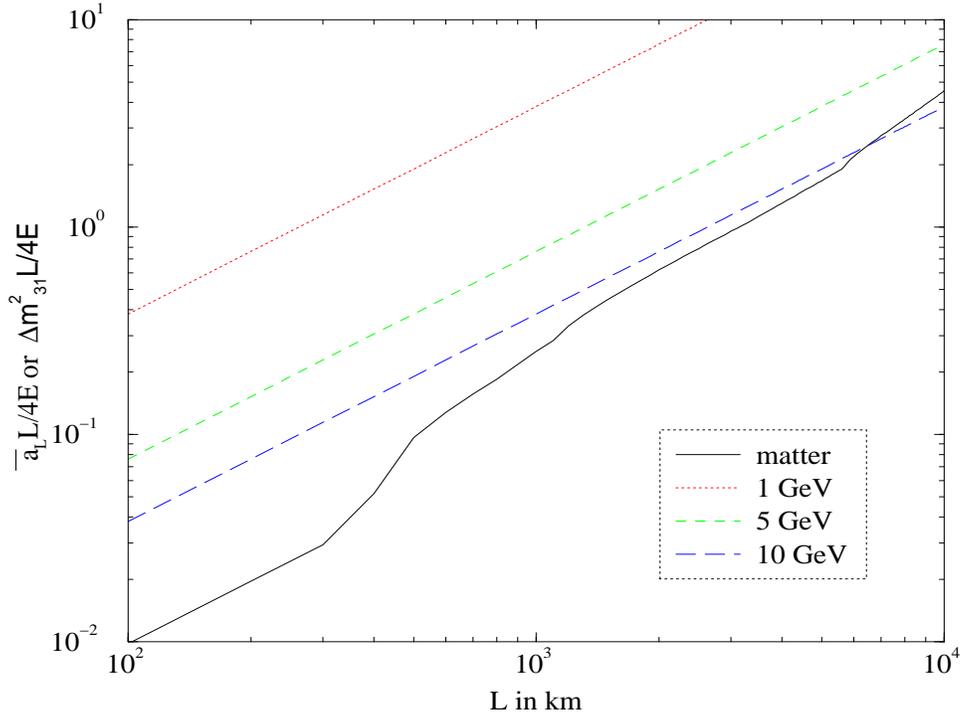,height=3.8in,width=5.in,angle=0}}
\caption{The comparison of the matter term against the $\Delta m_{31}^2$ 
term as a function of the neutrino baseline and for different neutrino 
energies. The matter term is independent of energy and is shown by the 
black solid line.
}
\label{matt}
\end{figure}

To check our other approximation concerning the 
matter potential we present in Figure \ref{matt} the 
comparison of the strength of the matter term {\it vis-a-vis} the 
term involving $\Delta m_{31}^2$. The solid line in the 
left-hand panel shows $\overline{a_L}L/4E$, where 
$\overline{a_L}$ is the ``average'' matter potential for a given 
neutrino baseline $L$ defined as 
\be
\overline{a_L} = \frac{1}{L}\int_0^L a(s,E) ds
\ee
For each experimental baseline $L$, $\overline{a_L}$ 
is computed using the PREM model \cite{ref10}.
Since $a(s,E)$ goes linearly with the neutrino energy $E$ (cf. Eq.(2)), the 
matter potential term $\overline{a_L}L/4E$ is independent of the 
energy of the neutrino 
beam. It is only a function of the average matter density and hence of 
the baseline $L$.
%We find that $a(L)$ increases almost linearly with 
%except for a small non-linearity around 
The dotted, dashed and long-dashed lines show 
$\Delta m_{31}^2 L/4E$ for $E=1$, 5 and 10 GeV respectively.
We note that for neutrinos with $E=3$ GeV the approximation 
given in Eq. (\ref{eq01}) works very well for all baselines shown. 
However for neutrino energies equal to or in excess of  
10 GeV our perturbative expansion 
for the matter term works only for smaller $L$ and 
breaks down at higher baselines.

\begin{figure}[t]
\centerline{\epsfig{figure=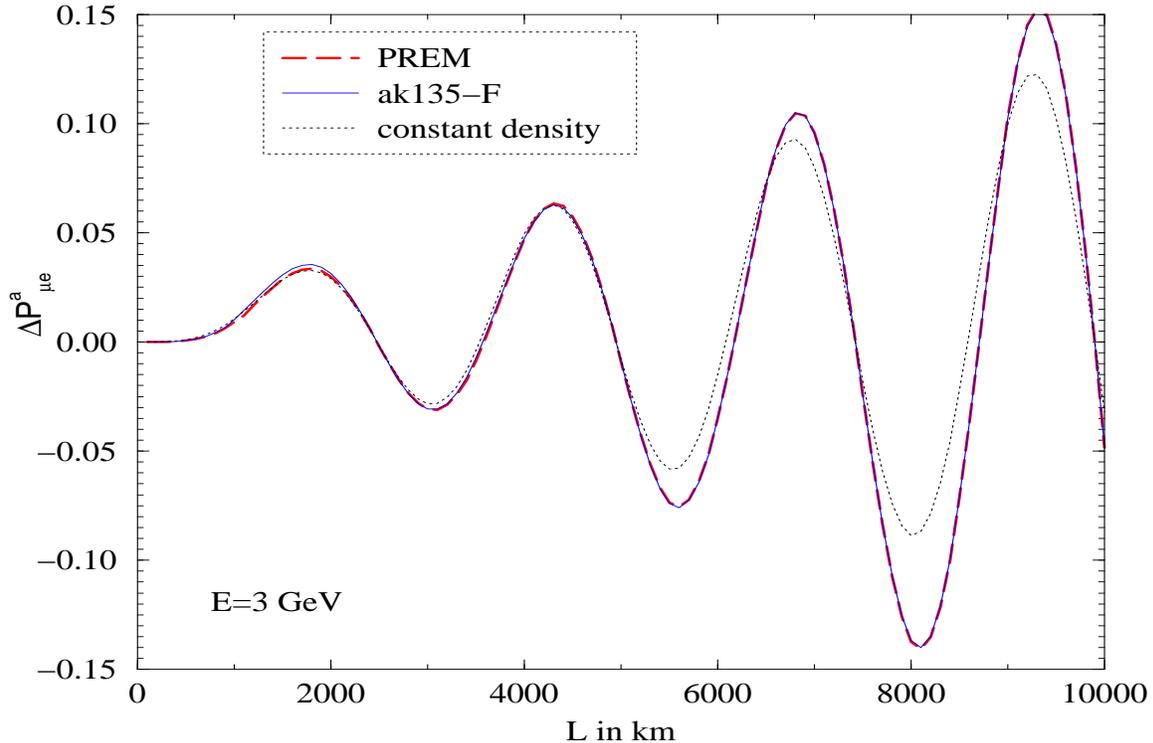,height=4.in,width=6.in}}
\caption{The CP asymmetry induced by matter effect ($\Delta P^a$) 
as a function of the baseline $L$ for the $\numu \rightarrow \nue$ 
oscillation 
channel with $E=3$ GeV. The dashed line 
and the solid line give the 
$\Delta P^a$ calculated using the PREM model and the ak135-F model 
respectively, 
while the dotted line is  
for a constant density earth with $\rho=3.28$ gm/cc. We have taken 
$\Delta m_{31}^2=3\times 10^{-3}$ eV$^2$, 
$\Delta m_{21}^2=7\times 10^{-5}$ eV$^2$,
$|U_{\mu 3}|^2=0.5$ and $|U_{e3}|^2=0.01$.
}
\label{delta}
\end{figure}

In Table 1 we give a summary of the magnitude of $\Delta P^a$ for
some specific cases. 
In Figure \ref{delta} we show $\Delta P^a$ as a function of the
baseline length $L$ for the $\nu_\mu \rightarrow \nu_e$ transition 
and with $E = 3$ GeV. 
We have checked that very similar behaviour is exhibited by 
$\Delta P^a$ for the transitions $\nu_e \rightarrow \nu_\tau$ and 
$\nu_\mu \rightarrow \nu_\tau$.
The solid and the dashed lines show the  $\Delta P^a$ 
obtained using our formalism for a varying density matter profile for the 
earth. The dashed line is for the PREM model \cite{ref10} while the solid line 
corresponds to the ak135-F model \cite{ak135f} for the earth matter profile.
The dotted line gives the corresponding values for 
a constant density earth with $\rho=3.28$ gm/cc.
The values of parameters used for generating the
figure and the table are shown in the captions.
Comparison of the solid and/or dashed lines with the constant density 
dotted line shows that the variation in the density profile can lead to 
a change in the oscillation probability. Even at $L=3000$ km, we note 
about 10\% difference in $\Delta P^a$ between the constant density case 
and the case corresponding to PREM or ak135-F. At smaller $L$ we note a 
small difference between $\Delta P^a$ corresponding to PREM and 
ak135-F, which just reflects the fact that ak135-F has more density 
fluctuation at smaller $L$ than PREM\footnote{It has been pointed 
out in \cite{Shan:2003vh} that even the so called realistic earth models 
like PREM and ak135-F neglect the {\it local} density fluctuations which can 
have an impact on the final oscillation probability and hence on the CP 
sensitivity of a specific experiment. However our expressions for 
the oscillation probabilities in varying density 
matter are completely general and 
can be applied to any earth matter density profile, local
variations notwithstanding.}. 
We see that the effect of density variation begins to be significant for $L >
5000$ kms. A caveat is that out formulae are not quantitatively reliable if 
$L$ and $E$ are such that either (\ref{eq01}) or the first of
conditions (\ref{eq02}) breaks down. (The second condition of
(\ref{eq02}) is always safely obeyed). Such is evidently not the case
in most of Figure 1.  Work is in progress to improve upon the first
order perturbation theory in $a(s,E)$ so that the dependence on these
conditions is reduced.

       \begin{table}
       \begin{center} 
       \begin{tabular}{|c|c|c|c|c|c|c|}
      \hline
      E/GeV & $U_{e3}$ & $|\Delta P^a_{\tau \mu}(732)|$ & 
      $|\Delta P^a_{e \mu}(732)|$ & $|\Delta P^a_{\mu e}(732)|$ &
      $|\Delta P^a_{\tau e}(732)|$ & $R$ in GeV
       \\
       \hline 
     1 & $0.001$ &$5.32  \times 10^{-7}$ &
                $5.32 \times 10^{-7}$ &
                $5.32 \times 10^{-7}$ &
                $5.32\times 10^{-7}$ & -1.54 \\
      1 & $0.01$ & $ 5.32\times 10^{-5}$ &
               $5.32 \times 10^{-5}$ &
               $5.32 \times 10^{-5}$ &
               $5.32 \times 10^{-5}$ & -1.54 \\
      1 & $0.1$ & $5.21 \times 10^{-3}$ &
              $ 5.21 \times 10^{-3}$ &
              $ 5.21 \times 10^{-3}$ &
              $ 5.11 \times 10^{-3}$ & -1.54 \\
      \hline
      E/GeV & $U_{e3}$ & $|\Delta P^a_{\tau \mu}(2500)|$ & 
      $|\Delta P^a_{e \mu}(2500)|$ & $|\Delta P^a_{\mu e}(2500)|$ &
      $|\Delta P^a_{\tau e}(2500)|$ & $R$ in GeV
      \\
      \hline
      3 & $0.001$ &$2.16  \times 10^{-7}$ &
                $2.16 \times 10^{-7}$ &
                $2.16 \times 10^{-7}$ &
                $ 2.16\times 10^{-7}$ & -19.42 \\
      3 & $0.01$ & $ 2.16\times 10^{-5}$ &
               $2.16 \times 10^{-5}$ &
               $2.16 \times 10^{-5}$ &
               $2.16 \times 10^{-5}$ & -19.42 \\
      3 & $0.1$ & $2.12 \times 10^{-3}$ &
              $ 2.12 \times 10^{-3}$ &
              $ 2.12 \times 10^{-3}$ & 
              $ 2.08 \times 10^{-3}$ & -19.42 \\
      \hline
      E/GeV & $U_{e3}$ & $|\Delta P^a_{\tau \mu}(3000)|$ & 
      $|\Delta P^a_{e \mu}(3000)|$ & $|\Delta P^a_{\mu e}(3000)|$ &
      $|\Delta P^a_{\tau e}(3000)|$ & $R$ in GeV
       \\
       \hline 
      5 & $0.001$ & $5.72 \times 10^{-6}$ &
                $5.72 \times 10^{-6}$ &
                $5.72 \times 10^{-6}$ &
                $5.72 \times 10^{-6}$ & -37.91 \\
      5 & $0.01$ & $5.72 \times 10^{-4}$ &
               $5.72 \times 10^{-4}$ &
               $5.72 \times 10^{-4}$ &
               $5.72 \times 10^{-4}$ & -37.91 \\
      5 & $0.1$ & $ 5.61 \times 10^{-2}$ &
              $ 5.61 \times 10^{-2}$ &
              $5.61  \times 10^{-2}$ & 
              $ 5.50 \times 10^{-2}$ & -37.91 \\
      \hline
      E/GeV & $U_{e3}$ & $|\Delta P^a_{\tau \mu}(3500)|$ & 
      $|\Delta P^a_{e \mu}(3500)|$ & $|\Delta P^a_{\mu e}(3500)|$ &
      $|\Delta P^a_{\tau e}(3500)|$ & $R$ in GeV \\
      \hline
      7 & $0.001$ & $7.03 \times 10^{-6}$ &
                $7.03 \times 10^{-6}$ &
                $7.03 \times 10^{-6}$ &
                $7.03 \times 10^{-6}$ & -52.10 \\
      7 & $0.01$ & $7.03 \times 10^{-4}$ &
               $7.03 \times 10^{-4}$ &
               $7.03 \times 10^{-4}$ &
               $7.03 \times 10^{-4}$ & -52.10 \\
      7 & $0.1$ & $6.89 \times 10^{-2}$ &
              $ 6.89 \times 10^{-2}$ &
              $ 6.89\times 10^{-2}$ & 
              $ 6.75 \times 10^{-2}$ & -52.10 \\
      \hline
       \end{tabular}
       \end{center}
      \caption{The magnitude of $\Delta P^a$ for some specific values of 
$E$ and $L$ and for the different oscillation channels. 
Values of $L$ in kms are shown in parantheses after
$\Delta P^a$.
Also shown is 
the value of the function $R(L,E)$ defined in Eq.(32). We have taken 
$\Delta m_{31}^2=3\times 10^{-3}$ eV$^2$, 
$\Delta m_{21}^2=7\times 10^{-5}$ eV$^2$ and 
$|U_{\mu 3}|^2=0.5$.
We present results for three different 
values of $|U_{e3}|$.
%We have used ${L/KM \over E/GeV} \sim O(10^2-10^3)$.
} 
      \label{tab}
      \end{table}

\section{Conclusions}

With the confirmation of neutrino flavour oscillations both in the 
atmospheric as well as the solar 
neutrino sectors, the focus now has shifted 
to the precise determination of the oscillation parameters 
involved. This will be possible in the currently planned and future long 
and very long baseline experiments involving conventional superbeams and 
neutrino factories. The determination of CP violation in the lepton 
sector and measurement of the CP phase will be the most interesting 
as well as the most challenging goal of these experiments. 
The matter effect induces a
``fake'' CP asymmetry
%Matter effects induce a CP asymmetry in the neutrino beam, 
even if there is no intrinsic CP violation in the neutrino sector. 
Thus a knowledge of
matter effect is extremely important in all neutrino CP violation 
studies. The matter effect also helps in ascertaining the sign of 
the atmospheric neutrino mass squared difference $\Delta m^2_{31}$
which has a bearing on the neutrino mass hierarchy with
profound theoretical and phenomenological implications. The effect
of earth matter on the survival and transition probabilities were 
studied earlier and ways to disentangle the ``real'' CP from the 
``fake'' CP due to matter were discussed. However most of these 
studies assumed a constant density of the earth matter.

In this paper we have used the evolution operator formalism to derive
the most general lowest order perturbative expressions for neutrino
flavour survival and transition probabilities in varying density
matter. We have worked in a perturbative scheme, where the oscillation
driven by $\Delta m_{31}^2$ is assumed to be much larger than those
driven by $\Delta m_{21}^2$ and the matter potential $a(s,E)$. We have
shown to the lowest nontrivial order that for a maximal mixing in the
$\numu-\nutau$ sector, the matter effect in the survival probability
$P_{\mu\mu}$ vanishes identically.  We have made numerical checks for
the validity of our perturbation approximation and conclude that as
long as $E$ $\lsim$ $10$ GeV, our approximation holds, at least upto a
baseline of $L\sim 4000$ km.  Finally, we have compared the results
obtained with the PREM and ak135-F density profiles with that for a
constant density earth matter.  Work is in progress to extend the
range of validity of our theory to longer baselines.

\bigskip
\medskip

\noindent{\bf Acknowledgements}

\medskip

B.B. acknowledges hospitality of Department of Theoretical
Physics, TIFR and of Theory Division, CERN.
S.C. thanks S.T. Petcov for discussions.
P.R. is indebted to Theory Group, SLAC and to
Santa Cruz Institute for Particle Physics, UCSC, for their hospitality.


\begin{thebibliography}{99}


\bibitem{ref1}
N. Cabibbo, Phys. Lett. {\bf B72}, 333 (1978);
V. D. Barger, K. Whistant and R. J. N. Philips, Phys. Rev. Lett. 
{\bf 45} 2084 (1980); S. Pakvasa, {\it Proc. XXth Intl. Conf. on
High Energy Physics}, Madison (1980), p 1164; S. M. Bilenky and
F. Niedermeyer, Sov. J. Nucl. Phys. {\bf 34}, 606 (1981)
[Yad. Fiz. {\bf 34}, 1091 (1981)].

\bibitem{ref2}
%\cite{Albright:2000xi}
%\bibitem{Albright:2000xi}
A. de Rujula, M.B. Gavela and P. Hernandez, Nucl. Phys. {\bf B547}, 21
(1999) [arXiv: hep-ph/9911390]. 
C.~Albright {\it et al.},
%``Physics at a neutrino factory,''
arXiv:hep-ex/0008064.
%%CITATION = HEP-EX 0008064;%%
%
%\cite{Apollonio:2002en}
%\bibitem{Apollonio:2002en}
M.~Apollonio {\it et al.},
%``Oscillation physics with a neutrino factory,''
arXiv:hep-ph/0210192.
%%CITATION = HEP-PH 0210192;%%
%
%\cite{Itow:2001ee}
%\bibitem{Itow:2001ee}
Y.~Itow {\it et al.},
%``The JHF-Kamioka neutrino project,''
arXiv:hep-ex/0106019.
%%CITATION = HEP-EX 0106019;%%
%
%\cite{Ayres:2002nm}
%\bibitem{Ayres:2002nm}
D.~Ayres {\it et al.},
%``Letter of intent to build an off-axis detector to study nu/mu $\to$ nu/e  oscillations with the NuMI neutrino beam,''
arXiv:hep-ex/0210005.
%%CITATION = HEP-EX 0210005;%%
%
%\cite{Lindner:2002vt}
%\bibitem{Lindner:2002vt}
M.~Lindner,
%``The physics potential of future long baseline neutrino oscillation  experiments,''
arXiv:hep-ph/0209083;
%%CITATION = HEP-PH 0209083;%%
%
%\cite{Dick:1999ed}
%\bibitem{Dick:1999ed}
K.~Dick, M.~Freund, M.~Lindner and A.~Romanino,
%``CP-violation in neutrino oscillations,''
Nucl.\ Phys.\ B {\bf 562}, 29 (1999)
[arXiv:hep-ph/9903308];
%%CITATION = HEP-PH 9903308;%%
%
%\cite{Cervera:2000kp}
%\bibitem{Cervera:2000kp}
A.~Cervera, A.~Donini, M.~B.~Gavela, J.~J.~Gomez Cadenas, P.~Hernandez, O.~Mena and S.~Rigolin,
%``Golden measurements at a neutrino factory,''
Nucl.\ Phys.\ B {\bf 579}, 17 (2000)
[Erratum-ibid.\ B {\bf 593}, 731 (2001)]
[arXiv:hep-ph/0002108];
%%CITATION = HEP-PH 0002108;%%
%
%\cite{Barger:2001yr}
%\bibitem{Barger:2001yr}
V.~Barger, D.~Marfatia and K.~Whisnant,
%``Breaking eight-fold degeneracies in neutrino CP violation, mixing, and  mass hierarchy,''
Phys.\ Rev.\ D {\bf 65}, 073023 (2002)
[arXiv:hep-ph/0112119]
%%CITATION = HEP-PH 0112119;%%
%
%\cite{Geer:2002pv}
%\bibitem{Geer:2002pv}
S.~Geer,
%``Neutrino factories: Physics potential,''
arXiv:hep-ph/0210113;
%%CITATION = HEP-PH 0210113;%%
%
%\cite{Yasuda:2002ii}
%\bibitem{Yasuda:2002ii}
O.~Yasuda,
%``Neutrino factories: Physics potential and present status,''
arXiv:hep-ph/0209127.
%%CITATION = HEP-PH 0209127;%%
%

\bibitem{ref3}
M. Fukugita and T. Yanagida, Phys. Lett. {\bf B175}, 45 (1986);
G. Branco, T. Morozumi, B. Nobre and M. N. Rebelo, Nucl. Phys. {\bf B617},
475 (2001); A. Joshipura, E. Pascos and W. Rodejohann, JHEP {\bf 0108},
029 (2001); W. Buchm\"uller and D. Wyler, Phys. Lett. {\bf B521}, 291
(2001); T. Endoh, T. Morozumi and A. Purwanto, Nucl. Phys. Proc. Suppl.
{\bf 111}, 291 (2002); J. Ellis amd M. Raidal, Nucl. Phys. {\bf B 643},
229 (2002); S. Davidson and A. Ibarra, arXiv: hep-ph/0206304;
G. C. Branco, R. Gonzalez Felipe, F. R. Joaquim and M. N. Rebelo, Nucl.
Phys. Proc. Suppl. {\bf 111}, 303 (2002); P. Frampton, S. Glashow
and T. Yanagida, arXiv: hep-ph/0208157; T. Endoh, S. Kaneko, S. K. Kang,
T. Morozumi and M. Tanimoto, Phys. Rev. Lett. {\bf 89} 231 601
(2002). 

\bibitem{ref4}
%\cite{Ahmad:2002jz}
%\bibitem{Ahmad:2002jz}
Q.~R.~Ahmad {\it et al.}  [SNO Collaboration],
%``Direct evidence for neutrino flavor transformation from neutral-current interactions in the Sudbury Neutrino Observatory,''
Phys.\ Rev.\ Lett.\  {\bf 89}, 011301 (2002)
[arXiv:nucl-ex/0204008];
%%CITATION = NUCL-EX 0204008;%%
%
%\cite{Ahmad:2002ka}
%\bibitem{Ahmad:2002ka}
Q.~R.~Ahmad {\it et al.}  [SNO Collaboration],
%``Measurement of day and night neutrino energy spectra at SNO and  constraints on neutrino mixing parameters,''
Phys.\ Rev.\ Lett.\  {\bf 89}, 011302 (2002)
[arXiv:nucl-ex/0204009];
%%CITATION = NUCL-EX 0204009;%%
%
%\cite{Fukuda:2002pe}
%\bibitem{Fukuda:2002pe}
S.~Fukuda {\it et al.}  [Super-Kamiokande Collaboration],
%``Determination of solar neutrino oscillation parameters using 1496 days  of Su per-Kamiokande-I data,''
Phys.\ Lett.\ B {\bf 539}, 179 (2002)
[arXiv:hep-ex/0205075];
%%CITATION = HEP-EX 0205075;%%
%
%\cite{Eguchi:2002dm}
%\bibitem{Eguchi:2002dm}
K.~Eguchi {\it et al.}  [KamLAND Collaboration],
%``First results from KamLAND: Evidence for reactor anti-neutrino  disappearance,''
Phys.\ Rev.\ Lett.\  {\bf 90}, 021802 (2003)
[arXiv:hep-ex/0212021].
%%CITATION = HEP-EX 0212021;%%

\bibitem{klpaper}
%\cite{Bandyopadhyay:2002en}
%\bibitem{Bandyopadhyay:2002en}
A.~Bandyopadhyay, S.~Choubey, R.~Gandhi, S.~Goswami and D.~P.~Roy,
%``The solar neutrino problem after the first results from KamLAND,''
arXiv:hep-ph/0212146;
%%CITATION = HEP-PH 0212146;%%
%
%\cite{Bandyopadhyay:2003du}
%\bibitem{Bandyopadhyay:2003du}
A.~Bandyopadhyay, S.~Choubey and S.~Goswami,
%``Exploring the sensitivity of current and future experiments to  Theta(odot),''
arXiv:hep-ph/0302243;
%%CITATION = HEP-PH 0302243;%%
%\cite{Barger:2002at}
%\bibitem{Barger:2002at}
V.~Barger and D.~Marfatia,
%``KamLAND and solar neutrino data eliminate the LOW solution,''
arXiv:hep-ph/0212126;
%%CITATION = HEP-PH 0212126;%%
%
%\cite{Fogli:2002au}
%\bibitem{Fogli:2002au}
G.~L.~Fogli, E.~Lisi, A.~Marrone, D.~Montanino, A.~Palazzo and A.~M.~Rotunno,
%``Solar neutrino oscillation parameters after first KamLAND results,''
arXiv:hep-ph/0212127;
%%CITATION = HEP-PH 0212127;%%
%
%\cite{Maltoni:2002aw}
%\bibitem{Maltoni:2002aw}
M.~Maltoni, T.~Schwetz and J.~W.~Valle,
%``Combining first KamLAND results with solar neutrino data,''
arXiv:hep-ph/0212129;
%%CITATION = HEP-PH 0212129;%%
%
%\cite{Bahcall:2002ij}
%\bibitem{Bahcall:2002ij}
J.~N.~Bahcall, M.~C.~Gonzalez-Garcia and C.~Pena-Garay,
%``Solar neutrinos before and after KamLAND,''
arXiv:hep-ph/0212147;
%%CITATION = HEP-PH 0212147;%%
%
%\cite{deHolanda:2002iv}
%\bibitem{deHolanda:2002iv}
P.~C.~de Holanda and A.~Y.~Smirnov,
%``LMA MSW solution of the solar neutrino problem and first KamLAND  results,''
arXiv:hep-ph/0212270.
%%CITATION = HEP-PH 0212270;%%
%
%\cite{Nunokawa:2002mq}
%\bibitem{Nunokawa:2002mq}
%H.~Nunokawa, W.~J.~Teves and R.~Zukanovich Funchal,
%``Determining the oscillation parameters by solar neutrinos and KamLAND,''
%arXiv:hep-ph/0212202;
%%CITATION = HEP-PH 0212202;%%
%
%\cite{Aliani:2002na}
%\bibitem{Aliani:2002na}
%P.~Aliani, V.~Antonelli, M.~Picariello and E.~Torrente-Lujan,
%``Neutrino mass parameters from Kamland, SNO and other solar evidence,''
%arXiv:hep-ph/0212212;
%%CITATION = HEP-PH 0212212;%%
%
%\cite{Balantekin:2003dc}
%\bibitem{Balantekin:2003dc}
%A.~B.~Balantekin and H.~Yuksel,
%``Global analysis of solar neutrino and KamLAND data,''
%arXiv:hep-ph/0301072;
%%CITATION = HEP-PH 0301072;%%
%
%\cite{Creminelli:2001ij}
%\bibitem{Creminelli:2001ij}
%P.~Creminelli, G.~Signorelli and A.~Strumia,
%``Frequentist analyses of solar neutrino data,''
%JHEP {\bf 0105}, 052 (2001)
%[arXiv:hep-ph/0102234 (v4)].
%%CITATION = HEP-PH 0102234;%%



\bibitem{ref5}
Y. Fukuda et al., Phys. Rev. Lett. {\bf 81}, 1562 (1998);
{\it ibid} {\bf 85}, 3999 (2001); T. Kajita and Y. Totsuka,
Rev. Mod. Phys. {\bf 73}, 85 (2001).

\bibitem{ref6}
M. Applonio et al., Phys. Lett. {\bf B466}, 415 (1999); F. Boehm
et al., Phys. rev. {\bf D64}, 112001 (2001).

\bibitem{ref7}
J. J. Gomez-Cadenas et al., arXiv: hep-ph/0105297; M. Akoi, K. Hagiwara
and N. Okamura, arXiv: hep-ph/0208223.

\bibitem{ref8}
I. Mocioiu and R. Shrock, Phys. Rev. {\bf D62}, 0153017 (2001);
%\cite{Mocioiu:2001jy}
%\bibitem{Mocioiu:2001jy}
I.~Mocioiu and R.~Shrock,
%``Neutrino oscillations with two Delta(m**2) scales,''
JHEP {\bf 0111}, 050 (2001);
%[arXiv:hep-ph/0106139];
%%CITATION = HEP-PH 0106139;%%
%
V. Barger, D. Marfatia and K. Whisnant, Phys. Rev. {\bf D65}, 073023 (2002);
%\cite{Ohlsson:2001et}
%\bibitem{Ohlsson:2001et}
T.~Ohlsson and H.~Snellman,
%``Neutrino oscillations with three flavors in matter of varying density,''
Eur.\ Phys.\ J.\ C {\bf 20}, 507 (2001).
%[arXiv:hep-ph/0103252].
%%CITATION = HEP-PH 0103252;%%

%\cite{Freund:1999gy}
\bibitem{Freund:1999gy}
M.~Freund, M.~Lindner, S.~T.~Petcov and A.~Romanino,
%``Testing matter effects in very long baseline neutrino oscillation  experiments,''
Nucl.\ Phys.\ B {\bf 578}, 27 (2000).
%[arXiv:hep-ph/9912457].
%%CITATION = HEP-PH 9912457;%%

\bibitem{ref9}
J. Arafune, M. Koike and J. Sato, Phys. Rev. {\bf D56} 3093 (1997);
{\it errtm. ibid.} {\bf D60} 119905 (1999). 

%\cite{Freund:1999vc}
\bibitem{Freund:1999vc}
M.~Freund and T.~Ohlsson,
%``Matter enhanced neutrino oscillations with a realistic earth density  profile,''
Mod.\ Phys.\ Lett.\ A {\bf 15}, 867 (2000).
%[arXiv:hep-ph/9909501].
%%CITATION = HEP-PH 9909501;%%

\bibitem{ref10}
A.M. Dziewonski and D.L. Anderson, Phys. Earth Planet Inter. {\bf 25}
297 (1981); S.V. Panasyuk, Reference Earth Model (REM) webpage,
{\it http://cfauves5.harvrd.edu/lana/rem/index.html}.

\bibitem{ref11}
M. Koike and J. Sato, Mod. Phys. Lett. {\bf A14} 1297 (1999); T. Ota
and J. Sato, Phys. Rev. {\bf D63} 093004 (2000);
%\cite{Jacobsson:2001zk}
%\bibitem{Jacobsson:2001zk}
B.~Jacobsson, T.~Ohlsson, H.~Snellman and W.~Winter,
%``Effects of random matter density fluctuations on the neutrino  oscillation transition probabilities in the earth,''
Phys.\ Lett.\ B {\bf 532}, 259 (2002);
%[arXiv:hep-ph/0112138].
%%CITATION = HEP-PH 0112138;%%
%
%\cite{Jacobsson:2002nb}
%\bibitem{Jacobsson:2002nb}
B.~Jacobsson, T.~Ohlsson, H.~Snellman and W.~Winter,
%``The effects of matter density uncertainties on neutrino oscillations in  the earth,''
arXiv:hep-ph/0209147.
%%CITATION = HEP-PH 0209147;%%

\bibitem{ref12}
T. Ota and J. Sato, arXiv: hep-ph/0211095.

%\cite{Shan:2001br}
\bibitem{Shan:2001br}
L.~Y.~Shan, B.~L.~Young and X.~m.~Zhang,
%``CP violating neutrino oscillation and uncertainties in earth matter  density,''
Phys.\ Rev.\ D {\bf 66}, 053012 (2002);
%[arXiv:hep-ph/0110414].
%%CITATION = HEP-PH 0110414;%%

%\cite{Shan:2003vh}
\bibitem{Shan:2003vh}
L.~Y.~Shan, Y.~F.~Wang, C.~G.~Yang, X.~Zhang, F.~T.~Liu and B.~L.~Young,
%``Modeling realistic Earth matter density for CP violation in neutrino oscillation,''
arXiv:hep-ph/0303112.
%%CITATION = HEP-PH 0303112;%%

\bibitem{ref13}
K. Kimura, A. Takamura and H. Yokomakura, arXiv: hep-ph/0203099. 

\bibitem {ak135f}
B.N.L. Kennett, E.R. Engdahl and R. Buland, Geophys. J. Int., 
{\bf 122}, 108, (1995); webpage
{\it http://wwwrses.anu.edu.au/seismology/ak135/intro.html}.

\end{thebibliography}
\end{document}